\documentclass[aps,twocolumn,showpacs]{revtex4}
\usepackage{epsf}
\usepackage{amssymb,graphicx}

\newcommand{\beq}{\begin{equation}}
\newcommand{\eeq}[1]{\label{#1} \end{equation}}

\newcommand{\bed}{\begin{displaymath}}
\newcommand{\eed}{\end{displaymath}}
\def\bea{\begin{eqnarray}}
\def\eea{\end{eqnarray}}
\begin{document}

\title{A critical view on the deeply bound $K^- pp$ system}

\author{V.K.~Magas$^{1}$, E. Oset$^{2}$, A. Ramos$^{1}$ and H. Toki$^3$}

\affiliation{
$^1$ Departament d'Estructura i Constituents de la Mat\`eria, \\
Universitat de Barcelona,
 Diagonal 647, 08028 Barcelona, Spain} \vskip 0.2cm
\affiliation{$^2$ Departamento de F\'{\i}sica Te\'orica and IFIC Centro Mixto\\
Universidad de Valencia-CSIC, Institutos de Investigaci\'on de Paterna \\
Apdo. correos 22085, 46071, Valencia, Spain} \vskip 0.2cm 
\affiliation{$^3$ Research Center for Nuclear Physics, Osaka University,\\
Ibaraki, Osaka 567-0047, Japan}

\begin{abstract}
We briefly review the situation around the claimed deeply bound
$K^-$ states in different recent experiments and concentrate
particularly on the state $K^- pp$ advocated by the  FINUDA
collaboration in nuclear $K^-$ absorption. We perform a
theoretical simulation of the process and show that the peak in
the $\Lambda p$ spectrum that was interpreted as a deep $K^- pp$
bound state corresponds mostly to the process $K^- p p \rightarrow
\Lambda p$ followed by final state interactions of the produced
particles with the daughter nucleus.
\end{abstract}


\date{\today}

\maketitle

\section{Introduction}

The theoretical predictions of deeply bound pionic atoms \cite{1,2,3,3b}, which
were observed in the $(d,~^3{\rm He})$ reaction \cite{4,5}, following a suggestion of the
theoretical studies of \cite{6,7}, have stimulated the search for bound states
of different hadrons in nuclei.  One of the natural continuations was the study
 of $K^-$ bound states. The large potential used at that time \cite{18} to
reproduce existing data on $K^-$ atoms fed the ideas that deeply
bound $K^-$ states (bound by $50-200$ MeV) could exist. The
question, of course, arises, as to whether the widths of the
states are narrow enough to distinguish between neighboring
states.  One should distinguish these deeply bound states from
other states bound by a few MeV in heavy nuclei that different
realistic potentials predict and which have not yet been observed
\cite{19,20,21,22}.

A step forward in the theoretical description of the $K^-$ nucleus
optical potential was given in \cite{29} establishing a link
between the optical potential and the presence of the
$\Lambda(1405)$ resonance, which, in free space, appears 27 MeV
below the $K^- p$ threshold. In the medium, the effect of Pauli
blocking on the excitation of intermediate $\bar{K}N$ states
demanded more energy to have the same phase space, as a
consequence of which the $\Lambda(1405)$ was produced at higher
energies. Correspondingly, the zero in the real part of the
in-medium $K^- p$ amplitude appears above the $K^- p$ threshold
and, as a consequence, the repulsive free-space $K^- p$ amplitude
at threshold turns into an attractive one in the medium.

The effect described above remains in all present theoretical
works as the driving mechanism to get an attractive $K^-$ nucleus
optical potential. Yet, the fact that the $K^-$ feels an
attractive self-energy in the medium facilitates the generation of
the $\Lambda(1405)$ at smaller energies, moving the zero of
 the amplitude back to
 smaller energies and, eventually,  ending up below the $K^- N $ threshold leading
 again to repulsion. It becomes clear at this point that a self-consistent
 calculation is needed, and it was done in \cite{30} where a small attraction
 was finally found with the resonance position barely modified.

Subsequent improvements introducing also the self-energy of the
pions and the baryons were done in \cite{33} by means of which a
potential was obtained that reproduced the data for the $K^-$
atoms \cite{21}. This potential, which automatically accounts for
new many body decay channels  with respect to former ones, like
$K^- NN \rightarrow \Sigma N, ~ \Lambda N, ~ \Sigma^* N$, leads to
deeply bound states with $B\approx 30$ MeV, but the width of the
states is around $100$ MeV \cite{21}. Potentials with the same
strength when self-consistency is imposed are found in
\cite{33,31,32}.

With this behavior for the ${\bar K}N$ dynamics in nuclei having
been established, the claim given in \cite{24} on the possible
existence of narrow deeply bound states in $A=3,4$ nuclei,
predicting a $A=3,\ I=0$ state with a binding energy of $100$ MeV,
was surprising. The situation became more puzzling when an
experiment searching for deeply bound $K^-$ states using the
reaction $^4$He(stopped $K^- , p)$  \cite{25}
found a signal
for a strange tri-baryon, $S^0(3115)$,  which was claimed not to
correspond to the state predicted in \cite{24}, since its
interpretation as a bound state would imply a binding energy
around $200$ MeV and the state had $I=1$. Subsequent corrections
of the potential in \cite{57} involving some spin-orbit effects
and relativistic corrections (which had been accounted for in
other works \cite{31,33}), plus some extra ad hoc modifications,
could lead to the extra binding of the kaon, as a consequence of
which the findings of \cite{25} have been lately presented as
evidence of deeply bound $K^-$ atoms \cite{26,satopanic}.

In a recent paper \cite{osettoki} two of us make a critical review
of the theoretical works \cite{24,57} and also of the
interpretation of the peak in the experiment of \cite{25}. The
main criticisms raised in \cite{osettoki} are the following. The
theoretical potential in \cite{24,57} eliminates the direct
coupling of $\pi \Sigma \rightarrow \pi \Sigma$ and $\pi \Lambda
\rightarrow \pi \Lambda$ in contradiction with the results of
chiral theory that establishes large couplings for them \cite{36}.
This assumption prevents the finding of two $\Lambda(1405)$ states
as appears recently in all chiral unitary approaches for the
$\bar{K}N$ system \cite{37,42,45,46,39} and which gets
experimental support from the study of the $K^- p \rightarrow
\pi^0 \pi^0 \Sigma^0$ reaction \cite{41} done in \cite{47}. The
experimental $\Lambda(1405)$ resonance, which is a superposition
of these two states, is assumed in \cite{24,57} to be a bound
state of $\bar{K}N$, contradicting the results of the chiral
theories and leading to a $\bar{K}N$ amplitude in $I=0$ twice as
large below the $K^- p$ threshold. The self-consistency shown to be
essential in \cite{30,33,31} is not implemented in \cite{24,57}
and, finally, the system is allowed to shrink to densities as
large as $10$ normal nuclear densities, $\rho=10 \rho_0$,
 at the center of the nucleus resulting
into a deeper $K^-$ nucleus potential. With all these assumptions
and approximations, the potential of \cite{57} has a strength of
about a factor twenty larger than that of the chiral approach.
Furthermore, the work of \cite{24,57} does not take into account
the many body decay channels considered in \cite{33} and, as was
shown in \cite{osettoki}, had they been included, and
with the densities claimed in \cite{24,57}, the width of the $K^-$
state should have been of the order of $200$ MeV or higher,
instead of the width of about $20$ MeV assumed in \cite{24,57}.

Convinced by these arguments that the peak  found at KEK \cite{25}
could not correspond to a deeply bound $K^-$ state, the authors of
\cite{osettoki} searched for other explanations and found a
natural mechanism that passed all tests. The peak is due to the
reaction $K^- NN \rightarrow \Sigma p$  in $^4$He leaving the
other nucleons as spectators. Such an interpretation demanded the
peak corresponding to the reaction $K^- NN \rightarrow \Lambda p$
be seen as well,
 which was indeed the case in the
experiment. This latter peak disappeared when a cut for ``fast"
pions was done (the pions coming from $\Lambda$ decay have
momentum in the range $61-196$ MeV/c) and was prominent in the
complementary cut of ``slow pions". The interpretation demanded
that the peaks had to be seen in other nuclei, a  test also passed
by the FINUDA data on the proton spectrum presented in \cite{finuda_new}. In addition, the
requirement that the daughter nucleus is just a spectator becomes
more difficult in heavier nuclei since the distortion of the
$\Lambda$ or $p$ particle in their way out through the nucleus
leads unavoidably
 to nuclear excitations. As a consequence of this, one expects the signals
  to fade gradually for heavier nuclei, which is also a feature of the FINUDA
  data on the proton spectrum.

With the situation of the KEK experiment clarified, the reanalysis of the
interpretation of the FINUDA data on the $\Lambda p$ invariant mass, leading to a claim of a bound state of
$K^- p p$ by $115$ MeV,  is an absolute necessity and this is the purpose
of the present work.

The FINUDA collaboration in \cite{finuda} looks for $\Lambda p$
events back to back following the $K^- p p$ absorption in $^6$Li,
$^7$Li and $^{12}$C nuclei. A narrow peak in the invariant mass is
observed, which is identified with $K^- p p \rightarrow \Lambda p$
removing just the binding energy. This implicitly assumes ground
state formation of the daughter nucleus since nuclear excitations
lead necessarily to a broad peak, namely the quasielastic peak. On
the other hand, a broad second peak of about $60 $ MeV width,
which is identified in \cite{finuda} as a signal for the $K^- p p$
bound state, is seen at smaller $\Lambda p$ invariant masses and
of much larger strength than the first peak. In the present work
we shall see that this peak, with its strength, position, width
and angular dependence corresponds to the same two-nucleon
absorption mechanism discussed above, but leading to nuclear
excitation, which has a much larger strength than the fraction
leading to the ground state in the daughter nucleus.

To reach the former conclusion, a computer simulation calculation is made
allowing the stopped kaons in the nucleus to be absorbed by a pair of nucleons
of a local Fermi sea. The nucleon and the $\Lambda$ emitted in the
$K^- p p \rightarrow \Lambda p $  and $K^- p n \rightarrow \Lambda n $
absorption processes are allowed to re-scatter with other nucleons in the nucleus
leading to nuclear breakup and producing a nucleon spectrum with a
distinct peak corresponding to one collision. This peak, which is analogous
to the quasielastic peak of any inclusive reaction like $(e,e')$,
$(\pi,\pi')$, $(p,p')$ etc, reproduces the experimental peak taken in
\cite{finuda} as a signal of the  $K^- p p$  bound state. Another peak,
broader and at smaller energies coming from baryon secondary collisions,
also appears both in our simulation and in the experimental data at the
same place and hence, an explanation for the whole experimental spectrum
is found, which does not require to invoke the creation of the
$K^- p p$ bound state.

\section{The theoretical framework}

The results of \cite{finuda} are presented for a mixture of data
from $^6$Li, $^7$Li and $^{12}$C. We shall perform our
calculations for these nuclei, as well as for the other two targets
installed in the
FINUDA apparatus, $^{27}$Al and $^{51}$V, although the invariant mass spectrum
was only shown for the three lighter nuclei. The reaction is
\beq K^- ~ A \rightarrow
\Lambda ~ p ~ A'
 \eeq{eq1}
  with stopped kaons. The $\Lambda p$
events are collected and, as done in the experiment, two cuts are
made: one selecting $p_\Lambda > 300$ MeV/c, to eliminate events
from $K^- p \rightarrow \Lambda \pi$, and another one imposing
$\cos \Theta_{\vec{p}_\Lambda \vec{p}_p}<-0.8$, to filter $\Lambda
p$ pairs going back to back.

The $K^-$ absorption at rest proceeds by capturing a slow $K^-$ in
a high atomic orbit of the nucleus, which later cascades down till
the $K^-$ reaches a low lying orbit, from where it is finally absorbed by the
nucleus.
According to Refs.~\cite{baer,gmitro}
from a study of the cascade of electromagnetic (EM) transitions after trapping
slow pions in nuclei, the nuclear absorption takes place from the
lowest atomic orbit and the next higher one where one has data for the atomic
energy shifts. This is so since, as soon as the absorption takes place, it
overcomes the EM transition precluding the observation of the X rays.
Application of the cascade absorption rates to
radiative pion capture has been done with success in
\cite{Roig:1985ge,Chiang:1989ni}.
The rational of the absorption probability from the $(n,l)$ levels is the
same for kaonic atoms and, hence, we also assume the absorption to take
place from the lowest level where the energy shift for atoms has been
measured, or, if it is not measured, from the level where the calculated
shift \cite{21} falls within the measurable range.
In the case of $^6$Li, $^7$Li, $^{12}$C, $^{27}$Al and $^{51}$V, the absorption
takes place from the orbits $2p$, $2p$, $2p$, $3d$ and $4f$ respectively. We
will also explore how the invariant $\Lambda p$ mass spectrum changes in
$^{12}$C when the $K^-$ is assumed to be
absorbed from the higher $3d$ orbit.

\begin{figure}
\vspace{-0.0cm}
\begin{center}
\includegraphics[width=8.6cm]{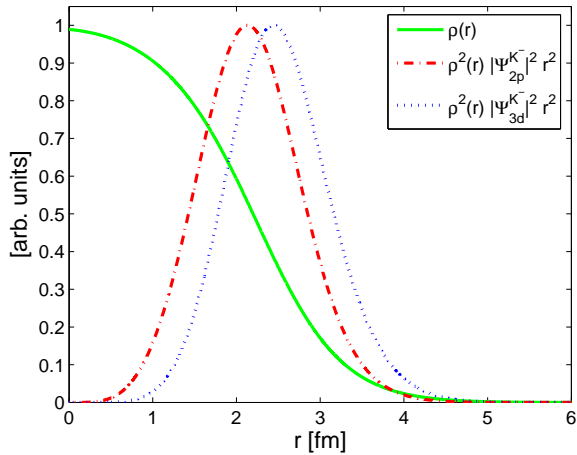}
\end{center}
\vspace{-0cm} \caption{(Color online) Nuclear density profile for $^{12}$C (solid line)
and overlap  $|\Psi_{K^-}(\vec{r}\,)|^2 r^2 \rho^2(r)$ for a $2p$ (dot-dashed line)
or $3d$ (dotted line) kaonic atom states.} 
\label{PsiK} \vspace{-0.0cm}
\end{figure}

The
cut $p_\Lambda > 300$ MeV/c in the experiment allows us to neglect
the one body $K^- N\rightarrow \Lambda \pi$ reaction and
concentrate exclusively on the two body absorption mechanism.
In Fig.~\ref{PsiK} we represent the density profile for $^{12}$C (solid line), 
together with 
the quantity $|\Psi_{K^-}(r)|^2 r^2 \rho^2(r)$, which represents an 
overlap function for the $K^-$ being absorbed by a pair of nucleons. The
dot-dashed line
represents the overlap when the $K^-$ is absorbed from an atomic $2p$ orbit, whereas
the dotted line represents the absorption from a $3d$ state. Both overlap
functions have been scaled such that their maximum value is 1, but we note that the
overlap for the $3d$ orbit is in fact 4 orders of magnitude smaller than that
for the $2p$ one.
However, this small magnitude does not mean that the absorption rate from
that state is so much smaller than that from the $2p$ state since in the
latter case one must also take into account the EM transition probability
from the $3d$ to $2p$ state in the cascade process \cite{baer,gmitro}. This is why we also
calculate the $\Lambda p$ invariant mass distribution following absorption
from the $3d$ state.

The $K^-$  absorption width from $p N$ pairs in a nucleus with $A$
nucleons is proportional in first approximation to the square of
the nuclear density \cite{ramos_94} multiplied by the probability
of finding the $K^-$ in the nucleus, hence, it is given by the
expression:
\begin{eqnarray}
\Gamma_A &=&\mbox{\rm Norm} \int d^3 \vec{r}\,
 |\Psi_{K^-}(\vec{r}\,)|^2 \left(\frac{\rho(r)}{2}\right)^2 \Gamma_m = \nonumber\\
&=& \mbox{\rm Norm} \int d^3 \vec{r}\, |\Psi_{K^-}(\vec{r}\,)|^2 \ \ 
2 \int \frac{d^3 \vec{p}_1}{(2\pi)^3} \Theta(k_F(r)-|\vec{p}_1|) \nonumber\\
& \times & 
2 \int \frac{d^3 \vec{p}_2}{(2\pi)^3} \Theta(k_F(r)-|\vec{p}_2|) \ \ 
 \Gamma_m(\vec{p}_1,\vec{p}_2,\vec{p}_K,\vec{r}\,)\ , 
\label{eq2a}
\end{eqnarray}
 where we have used
$\rho(r)=4 \int \frac{d^3 \vec{p}_1}{(2\pi)^3}
\Theta(k_F(r)-|\vec{p}_1|)$ for the nucleons with momenta
$\vec{p}_1$, $\vec{p}_2$ in the Fermi sea, $k_F$ is the Fermi
momentum at the local point, $\Theta$ is the step function,
$\mbox{\rm Norm}$ is a proportionality constant that has
dimensions of $R^6$ and
$\Gamma_m(\vec{p}_1,\vec{p}_2,\vec{p}_K,\vec{r})$ is the in-medium
decay width for the $K^- p N \rightarrow \Lambda N$ process
given by
\begin{eqnarray}
 \Gamma_m(\vec{p}_1,\vec{p}_2,\vec{p}_K,\vec{r}\,)= \int
 \frac{d^3 \vec{p}_\Lambda}{(2\pi)^3} \frac{M_\Lambda}{p_\Lambda^0}
 \int \frac{d^3 \vec{p}_N}{(2\pi)^3} \frac{M_N}{p_N^0}& (2\pi)^4 &
 \nonumber \\
  \times 
 \delta^4(p_\Lambda+p_N-p_1-p_2-p_K)  & &  \nonumber \\
  \times  \frac{ \overline{\sum_{s_i}}\sum_{s_f}|T_{K^- p N \to \Lambda N}|^2 }
{m_K+p_1^0+p_2^0}\Theta(|\vec{p}_N|-k_F(r)) & & \ ,
\label{eq2b}
 \end{eqnarray}
where $\vec{p}_K$, $\vec{p}_\Lambda$, $\vec{p}_N$ are momenta of the corresponding particles, and  
$\frac{ \overline{\sum_{s_i}}\sum_{s_f}|T_{K^- p N \to \Lambda N}|^2 }{m_K+p_1^0+p_2^0}$ will be
assumed to be approximately
 constant
for $\vec{p}_K=0$, and $|\vec{p}_{1}|,|\vec{p}_{2}| \ll M$.

Now we insert Eq.~(\ref{eq2b}) into Eq.~(\ref{eq2a}), perform the integration
over $d^3 \vec{p}_1$ and the angular part of $d^3 \vec{p}_2$, and absorb all the
constants into $\mbox{Norm}$:
\begin{eqnarray}
&&\Gamma_A = \mbox{\rm Norm} \int d^3 \vec{r} \, |\Psi_{K^-}(\vec{r})|^2
\int
\frac{d^3 \vec{p}_\Lambda} {p_\Lambda^0} \int \frac{d^3
\vec{p}_N}{p_N^0} \nonumber \\
&\times & \int dp_2 p_2
\frac{p_\Lambda^0+p_N^0-p_2^0-m_K}{|\vec{P}|} ~\Theta(1-A^2) \nonumber \\
&\times & \Theta\left(k_F(r)-\sqrt{|\vec{P}|^2+|\vec{p}_2|^2-2|\vec{P}||\vec{p}_2| A}\right) \nonumber \\
&\times & 
\Theta(k_F(r)-|\vec{p}_2|) \Theta(|\vec{p}_N|-k_F(r)) \ ,
\label{eq2c}
\end{eqnarray}
where $\vec{P}=\vec{p}_\Lambda+\vec{p}_N$, and $A$ provides the
cosine of the angle between $\vec P$ and $\vec p_N$,
\begin{eqnarray}
& A & \equiv
\cos \Theta_{\vec{P}\vec{p}_N} \equiv \nonumber \\
& \equiv & \frac{ M_N^2+\vec P^2+\vec
p\,^2_2-(p_\Lambda^0+p_N^0-p_2^0-m_K)^2 }{2|\vec P||\vec p_2|}\,.
\label{eq2d}
\end{eqnarray}
In order to take into account the propagation of the produced
nucleon and $\Lambda$ through the nucleus after $K^-$  absorption
we insert a kernel $K(\vec p, \vec r\,)$ into Eq.~(\ref{eq2c}) for
both particles. This kernel has been used in Monte Carlo
simulations of pion nucleus reactions \cite{montecarlo}, proton
nucleus reactions \cite{carrasco}, electron nucleus reactions
\cite{gil}, leading to good reproduction of inclusive cross
sections and selected channels of one and two nucleon emission,
etc. Thus the final expression for the $K^-$  absorption width
from $p N$ pairs in nucleus $A$ is given by
\begin{eqnarray}
&&\Gamma_A = \mbox{\rm Norm} \int d^3 \vec{r} \, |\Psi_{K^-}(\vec{r})|^2
\int
\frac{d^3 \vec{p}_\Lambda} {p_\Lambda^0} \int \frac{d^3
\vec{p}_N}{p_N^0} \nonumber \\
&\times & \int dp_2 p_2
\frac{p_\Lambda^0+p_N^0-p_2^0-m_K}{|\vec{P}|} ~\Theta(1-A^2) \nonumber \\
&\times & \Theta\left(k_F(r)-\sqrt{|\vec{P}|^2+|\vec{p}_2|^2-2|\vec{P}||\vec{p}_2| A}\right) \nonumber \\
&\times & 
\Theta(k_F(r)-|\vec{p}_2|) \Theta(|\vec{p}_N|-k_F(r))
 \nonumber \\
& \times & K(\vec p_\Lambda,\vec r\,) K(\vec p_N,\vec r\,)\ .
\label{eq2}
\end{eqnarray}

The procedure accomplished by Eq.~(\ref{eq2}) is the following. A kaon at
rest is absorbed from the surface of the nucleus by two nucleons
($pp$ or $pn$) whose momenta are chosen randomly from the local Fermi sea,
with a Fermi momentum $k_F(r)=\left( 3\pi^2 \rho(r)/2 \right)^{1/3}$,
 where $\rho(r)$ is the experimental nuclear density.
Energy and momentum conservation is demanded according to phase
space and hence two momenta of the emitted $p\ (n)$ and $\Lambda$
are generated. Then, the primary nucleon is allowed to re-scatter with
nucleons in the nucleus by means of
\begin{eqnarray}
pN &\rightarrow& p^\prime N^\prime \nonumber  \\
np &\rightarrow& pn\quad ({\rm fast}\ n\ {\rm to~ fast}\ p)
\label{eq3}
\end{eqnarray}
 according to a probability per unit length given by
$\sigma \rho(r)$, where $\sigma$ is the experimental $NN$ cross
section at the corresponding energy. The $\Lambda$ is allowed to
scatter similarly, assuming a cross section of $\sigma_{\Lambda
N}\equiv \frac{2}{3} (\sigma_{pN}+\sigma_{nN})$, which is a
reasonable choice based on the observation that realistic $YN$
potentials predict a $\Lambda$ mean-field potential depth in
nuclei of about 2/3 that of the nucleon \cite{nose}. The nucleons
from the Fermi sea that
 participate in the collisions are also chosen randomly, the variables of
 the scattering particles are boosted to their Center of Mass (c.m.) frame,
 an angular distribution is generated according to experiment
 (for $\Lambda N$ it is assumed to be the same as for $NN$) and a boost
 back to the Lab frame is performed. After several possible collisions, one or
 more nucleons
 and a $\Lambda$ emerge from the nucleus and the invariant mass of all possible
 $\Lambda p$ pairs, as well as their
 relative angle, are evaluated for each Monte Carlo event
  and stored conveniently to generate the required
 histograms.

\section{Formation of the ground state of the final daughter nucleus}

Absorption of $K^-$ from a nucleus leaving the final daughter nucleus in its
ground state gives rise to a narrow peak in the $\Lambda p$ invariant mass
distribution, as it is observed in the spectrum of
\cite{finuda}. Our local density formalism, in which the hole levels in the Fermi
sea form a continuum of states, cannot handle properly transitions to discrete
states of the daughter nucleus, in particular to the ground state.
For this reason, we will remove in our calculations those events in
which the $p$ and $\Lambda$ produced after $K^-$ absorption leave the nucleus
without having suffered a secondary collision.

A precise value for the strength going into the ground state
of the daughter nucleus would require a quantum mechanical calculation with a
good description of the nuclear wave functions. However, for the purpose of
finding an interpretation to 
the second peak of the FINUDA experiment of about 60 MeV wide, 
this knowledge is not strictly necessary.
In this section we give qualitative arguments to estimate
the strength of the ground-state to ground-state process, with the only aim
of showing that it represents a moderate fraction of the total, meaning that
a significant amount of strength will go
to nuclear excitations, generating the typical quasiparticle
peak which the FINUDA collaboration interprets as a bound $K^- pp$ state.
Let's take as an example the case of $^7{\rm Li}$.
The process under consideration would be
\beq
K^- ~^7{\rm Li} \rightarrow p ~\Lambda ~^5{\rm H} \,.
\eeq{eq4}

\begin{figure}[h]
\vspace{-0.0cm}
\begin{center}
\includegraphics[width=8.6cm]{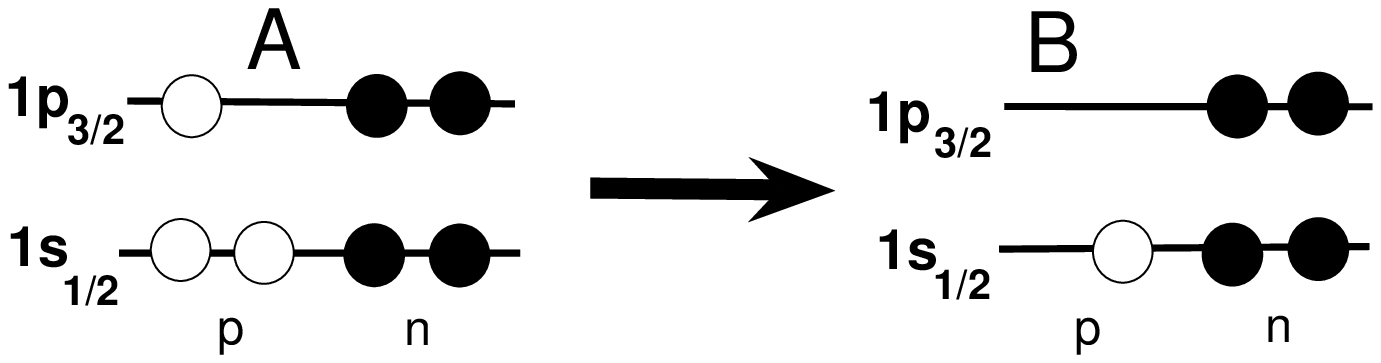}
\end{center}
\vspace{-0.0cm}
\caption{ A) The ground state of $^7$Li; B) The ground state of $^5$H. }
\label{fig1}
\vspace{-0.0cm}
\end{figure}

A simple shell model structure of $^7$Li is given in Fig.~\ref{fig1}A. 
In the same picture the ground state (g.s.) of the
daughter nucleus after the removal of two protons would be given
by Fig.~\ref{fig1}B. It is clear that even if $^5$H had an
approximate structure like in Fig.~\ref{fig1}B, its orbitals are
different from those of the $^7$Li, which means that the overlap
of the corresponding wave functions is smaller than $1$:
\beq |<
\Psi_{^5{\rm H}}(gs) |  \Psi_{^7{\rm Li}}(gs) (pp)^{-1} >| <1\,.
\eeq{eq5}

Actually, for the discussion that follows, we do not need to know
the precise value of this overlap. Values for it of the order of
$0.5-0.7$ are already large, which would lead to a formation
probability, defined as
\beq {\rm Form.\ prob.}=|< \Psi_{^5{\rm
H}}(gs) |  \Psi_{^7{\rm Li}}(gs) (pp)^{-1} >|^2\,,
\eeq{eq6}
 of the order or smaller than $0.3$. The formation probability is a
 concept used in $\alpha$ decay, where such an overlap plays an important role
 in the decay rates when four particles are removed from the father nucleus
 \cite{prestonbhaduri}.

\begin{figure}
\vspace{-0.0cm}
\begin{center}
\includegraphics[width=8cm]{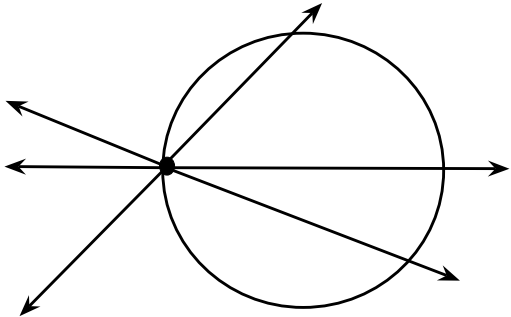}
\end{center}
\vspace{-0cm}
\caption{A simplified $K^-$ absorption picture for the estimation of the survival
 probability.}
\label{fig2}
\vspace{-0.0cm}
\end{figure}

The formation probability is not all what matters in the reaction.
In $\alpha$ decay one has to consider the probability of
penetrating through the potential barrier. In the present case the
formation probability has to be weighted by the ``survival
probability", corresponding to the situation in which both the $p$
and the $\Lambda$ cross the daughter nucleus
 without any collision. This is so because any such collision would excite
 the nucleus, which would then not remain in its ground state.
By looking at Fig. \ref{fig2} one observes that, for the $K^-$
being absorbed on the surface, only one baryon ($p$ or $\Lambda$)
will cross the nucleus, and the survival probability would be \beq
P_s={\rm e}^{-\int_0^\infty \sigma \rho(\vec{r}\,^\prime) dl }\,,
\eeq{eq7} where $\vec{r}\,^\prime=\vec{r}+l\vec{p}/p$ is the production point
 and $\sigma
\simeq 20$ mb for $581$ MeV/c, corresponding to the momentum of
the proton coming from $K^- pp \rightarrow p \Lambda $ absorption.
A simple estimate using Eq.~(\ref{eq7}) gives $P_s\simeq 0.4$ for
 $^{12}$C, which is confirmed by the
more accurate value extracted from our Monte Carlo
simulations, namely $P_s=0.397$ assuming $K^-$ absorption from the $2p$ orbit 
and $P_s=0.396$ for absorption from the $3d$ orbit. 
In the case of $^{7}$Li our Monte Carlo simulation gives a value of $P_s=0.56$,
while for other nuclei we obtain 
$P_s=0.59$ in $^{6}$Li, 
$P_s=0.26$ in $^{27}$Al, and 
$P_s=0.18$ in $^{51}$V.
Combining this factor with the formation probability
of $\leq$ 0.3, we finally estimate a probability smaller than about 15\%  
for having the daughter nucleus in its ground state after a $K^-pp$ absorption
event. Although crude, the former evaluation of the role of the
ground state formation is sufficient for our purposes, since we are only 
interested in the position and shape of the quasielastic peak, which
does not depend on the precise value of the
proportion of strength going to the ground state relatively to that going to
nuclear excitations.

It should also be mentioned here that the possibility of removing
two protons from the s-wave orbit in Fig. \ref{fig1} leading to a final
 $^5$H excited nucleus in the continuum without further interaction of the
 $p$ or $\Lambda$ is negligible due to the small overlap between these two states.
 This means in practice that the
 excitation of the nucleus will require the secondary collision of the $p$
 or $\Lambda$ after the $K^-pp$ absorption process.
This is indeed what happens, for instance, in $(p,p')$ collisions, where the
strength of the cross section to elastic or bound excited states is very
small compared to that of nuclear breakup producing the quasielastic peak.

In  Fig. 3 of Ref. \cite{finuda} a narrow peak is observed around
$\sqrt{s}=2350$ MeV, already interpreted by the authors as the
g.s. formation process described above. 
If this process accounts for a relatively small fraction of the $K^-$ absorption
rate, obviously the bulk of the strength will go into the region of
smaller invariant $\Lambda p$ mass. One can then immediately guess
that the broad peak to the left of the narrow peak in the spectrum
of Fig. 3 of \cite{finuda} will account for these events. To check
this hypothesis we present in the next section the distribution of
strength for the events in which the emitted particles following
$K^-$ absorption undergo collisions in their way out of the
daughter nucleus, after having applied the same cuts in the
$\Lambda$ momentum and the $\Lambda p$ relative angle as in the
experiment \cite{finuda}.

\section{Spectrum of the secondary collisions}

The spectrum of secondary collisions is generated from our
computer simulation algorithm of Eq.~(\ref{eq2}) which allows the
particles emitted after $K^-$ absorption to re-scatter in their
way out of the daughter nucleus. The nucleons move under the
influence of a mean field potential which we take to be given by
the Thomas-Fermi expression: 
\beq V(r)=-\frac{k_F(r)^2}{2m_N}\,.
\eeq{eq78} 
For each nucleus under consideration, an extra constant binding $\Delta$
is added to the hole nucleon spectrum, such that the maximum $\Lambda N$
invariant mass allowed by our model, $m_{K^-} + 2 M_p -2\Delta$,
corresponds to the actual invariant mass reached in $K^-$ absorption leading to
the ground state of the daughter nucleus, $m_{K^-}+ M(A,Z)-M(A-2,Z-2)$. Taking
the experimental values of the nuclear masses we obtain $\Delta=15.2$,
16.2, 12.8, 10.7 and 9.6 MeV for 
$^6$Li, $^7$Li, $^{12}$C, $^{27}$Al and $^{51}$V, respectively.

\begin{figure}
\vspace{-0.0cm}
\begin{center}
\includegraphics[width=8.6cm]{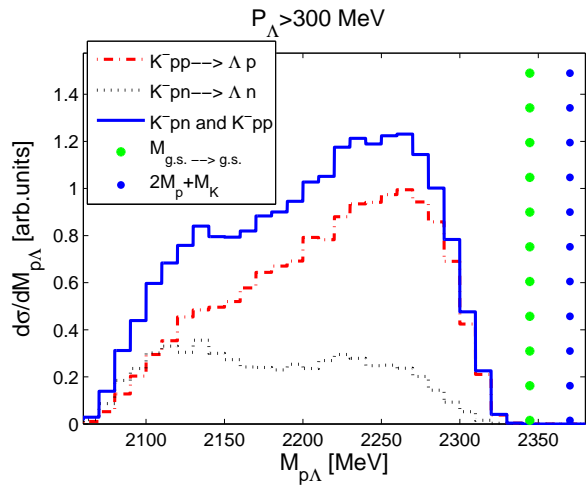}
\end{center}
\vspace{-0.0cm} \caption{(Color online) Invariant mass of $\Lambda p$
distribution for $K^-$ absorption in $^{12}$C with maximum one
collision of the outgoing particles with the daughter nucleus.}
\label{fig3} \vspace{-0.0cm}
\end{figure}

\begin{figure}
\vspace{-0.0cm}
\begin{center}
\includegraphics[width=8.6cm]{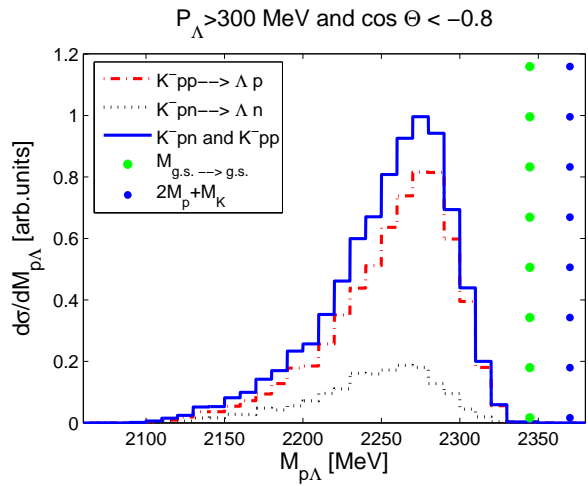}
\end{center}
\vspace{-0.0cm} \caption{(Color online) The same as Fig. \ref{fig3}, but imposing
the experimental angle cut for back to back events, $\cos
\Theta_{\vec{p}_\Lambda \vec{p}_p}<-0.8$. } \label{res1}
\vspace{-0.0cm}
\end{figure}

The results of the calculated spectra with just one collision
after $K^-$ absorption in $^{12}$C are shown in Fig. \ref{fig3}. A momentum cut
of $p_\Lambda > 300$ MeV/c has already been imposed since this condition is
fulfilled by almost all absorption events.  
On inspecting the thick line, which corresponds to the sum of 
the two possible absorption mechanisms, $K^-pp$ and $K^-pn$, with final state 
interactions (FSI), one can see a main
bump in the spectrum, which peaks around the same position and is
somewhat broader than the main peak shown in the insert of Fig. 3 of \cite{finuda}.
Since one measures the $\Lambda p$ invariant mass, the main contribution comes 
from $K^- p p \rightarrow \Lambda p$ absorption.
The contribution from the $K^- p n \rightarrow \Lambda n$ reaction
followed by $np\rightarrow pn$ is represented in the figure
by the dotted line. We can see that in the most interesting
region, $\sqrt{s}=2250-2300$ MeV, this process contributes only
by about $20$ \% , while in the lower invariant mass region it
becomes as important as the $K^- p p \rightarrow \Lambda p$
reaction. Low invariant
 masses correspond to the situation of $\Lambda p$ pairs 
 with a low energy proton
  kicked above Fermi sea in a collision with the primary nucleon. We will see below
   that the amount of such events will increase when we allow more than
   one secondary
   collision.

\begin{figure}
\vspace{-0.0cm}
\begin{center}
\includegraphics[width=8.6cm]{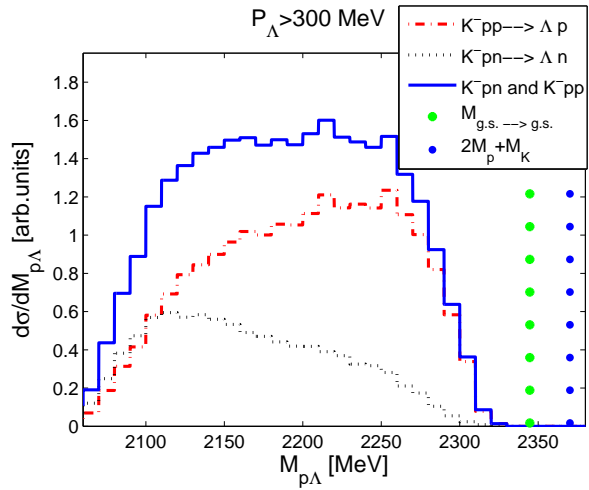}
\end{center}
\vspace{-0.0cm} \caption{(Color online) Invariant mass of $\Lambda p$
distribution for $K^-$ absorption in $^{12}$C allowing up to three
collisions of the outgoing particles with the daughter
 nucleus.}
\label{fig5}
\vspace{-0.0cm}
\end{figure}

\begin{figure}
\vspace{-0.0cm}
\begin{center}
\includegraphics[width=8.6cm]{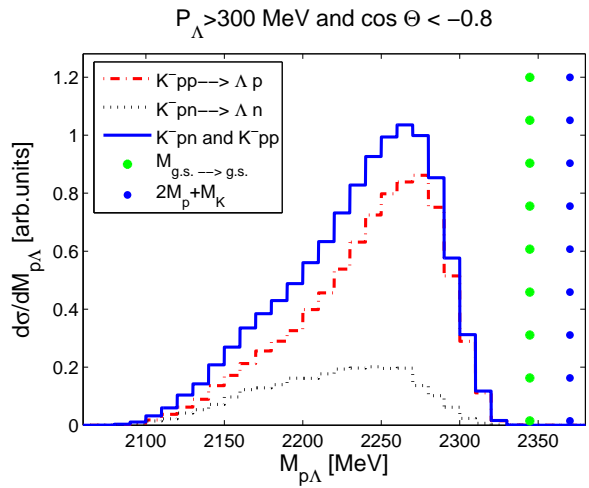}
\end{center}
\vspace{-0.0cm} \caption{(Color online) The same as Fig. \ref{fig5}, but imposing
the experimental angle cut for back to back events, $\cos
\Theta_{\vec{p}_\Lambda \vec{p}_p}<-0.8$. } \label{res2}
\vspace{-0.0cm}
\end{figure}

Next we impose the remaining experimental restriction in our calculated
spectrum, namely the back to back
angle cut, $\cos \Theta_{\vec{p}_\Lambda \vec{p}_p}<-0.8$. This has a
substantial effect, as can be seen in Fig. \ref{res1}, narrowing
considerably the invariant $\Lambda p$ mass spectrum, which looks
now much more similar to the one in Fig. 3 of \cite{finuda}.

We can go further by allowing more collisions of the proton or
$\Lambda$. In Fig. \ref{fig5} and \ref{res2} we show,
respectively, the spectrum of the $\Lambda p$ 
invariant mass
without and with the experimental angle cut, and allowing up to
three collisions of the  $\Lambda$ or $p$ particles with the
nucleons in the Fermi sea of the daughter nucleus.
With respect to the one-collision case shown in Fig.~\ref{res1}, 
the spectrum of Fig.~\ref{res2} shows a more clear peak at 2260 MeV, 
as in the insert of Fig. 3 in \cite{finuda}, and
develops additional strength in the lower energy region, also
present in the experimental data.
The level of agreement of the prediction with the experimental data is the
same as in other theoretical studies of the quasielastic peak 
\cite{montecarlo,carrasco,gil}.

\begin{figure}
\vspace{-0.0cm}
\begin{center}
\includegraphics[width=8.6cm]{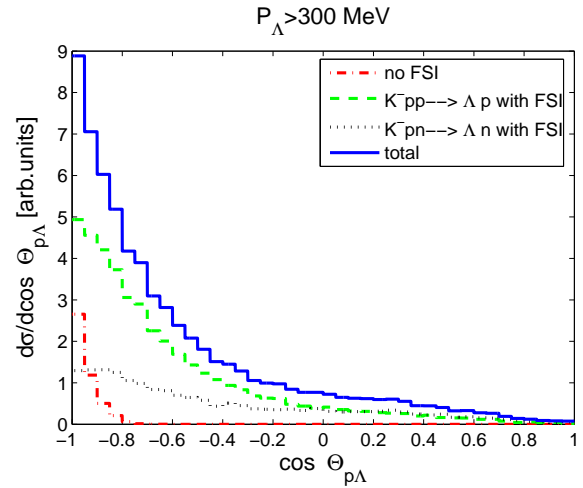}
\end{center}
\vspace{-0.0cm}
\caption{(Color online) Opening angle distribution between a $\Lambda$ and a 
proton in $^{12}$C, allowing up to three
collisions of the outgoing particles with the daughter
 nucleus.
 We assumed a 10\% probability for
 the no FSI events, which correspond to those leaving the daughter 
nucleus in its ground state after $K^-$ absorption. }
\label{fig4}
\vspace{-0.0cm}
\end{figure}

The angular distribution of the $\Lambda p$ events is shown in
Fig. \ref{fig4}, where we have added to the spectrum with
secondary collisions the calculated strength of the events corresponding to the
g.s. to g.s. transition, which has been imposed to be about 10\%. This fraction
is about the one shown in Fig.~3 of Ref.~\cite{finuda} and roughly agrees with
the magnitude estimated in previous section for $^7$Li.
A larger fraction than this would lead to an even more forward peaked
distribution. However, as we discussed above, all the strength of these
events is accumulated in a narrow peak of the $\Lambda p$ mass spectrum
around the line of the light circles of Figs.~\ref{res1},\ref{res2} and has 
no influence on the broad peak of the spectrum.
In our
simulations, the g.s. to g.s. reaction corresponds to the case of
no secondary collisions. We observe that the angular dependence of
the peak for a transition to the g.s. of the daughter nucleus is
obviously back to back, since the three particles participating in
the absorption mechanism are practically at rest. Even assuming
that the $K^- pp$ system gets about $200$ MeV/c momentum from the
Fermi motion of the protons, this only leads to $4$ MeV recoil
energy for the daughter nucleus, making the $\Lambda p$ system
move with velocity given as $\frac{200\ {\rm MeV}}{m_K+2M_p}$, and the angle
between $\Lambda$ and $p$ fulfills $\cos \Theta_{\vec{p}_\Lambda
\vec{p}_p}<-0.95$. Thus, all events of $K^-$ absorption
contributing to the narrow peak in the FINUDA experiment fall
within the experimental cuts.

As we can see, the angular distribution of the total strength in
Fig. \ref{fig4} peaks back to back as in the experiment. We note that one should only expect a qualitative agreement with
the angular distribution shown in Fig.~2 of \cite{finuda}, since that one is not
corrected for the detector acceptance.

\begin{figure}
\vspace{-0.0cm}
\begin{center}
\includegraphics[width=8.6cm]{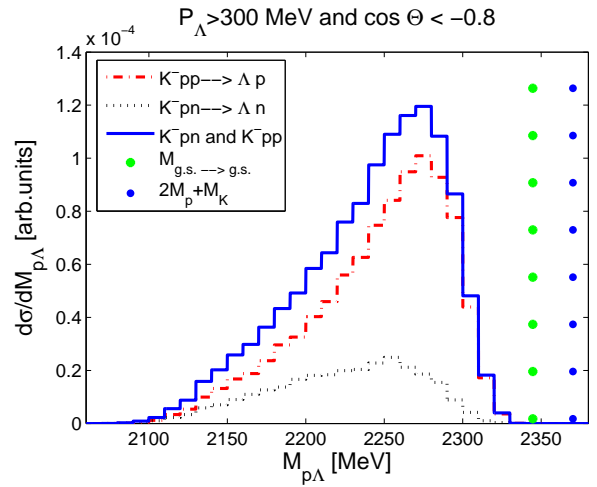}
\end{center}
\vspace{-0.0cm} \caption{(Color online) The same as Fig. \ref{res2}, but assuming that
$K^-$ absorption proceeds from the atomic $3d$ orbit.} \label{3d}
\vspace{-0.0cm}
\end{figure}

In order to explore the dependence of ours results on the
atomic orbit assumed for the $K^-$, we show in Fig.~\ref{3d} the 
calculated $\Lambda p$ invariant mass distribution for the $K^-$ being
absorbed in $^{12}$C from a higher $3d$ orbit. 
It is clearly seen that the shape and centroid of the distribution are
barely altered. Therefore, if the  absorption of the $K^-$  proceeds from an
arbitrary superposition of  the $2p$ and $3d$ orbits, the final spectrum
would look just as those shown in Figs. \ref{res2} and \ref{3d}.

\begin{figure}
\vspace{-0.0cm}
\begin{center}
\includegraphics[width=8.6cm]{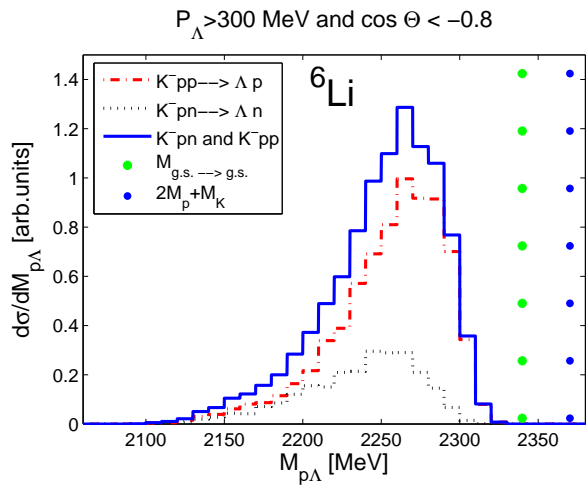}
\end{center}
\vspace{-0.0cm} \caption{(Color online) Invariant mass of $\Lambda p$
distribution for $K^-$ absorption in  $^{6}$Li and allowing up to two collisions of the outgoing particles
with the daughter nucleus.} \label{li6}
\vspace{-0.0cm}
\end{figure}

\begin{figure}
\vspace{-0.0cm}
\begin{center}
\includegraphics[width=8.6cm]{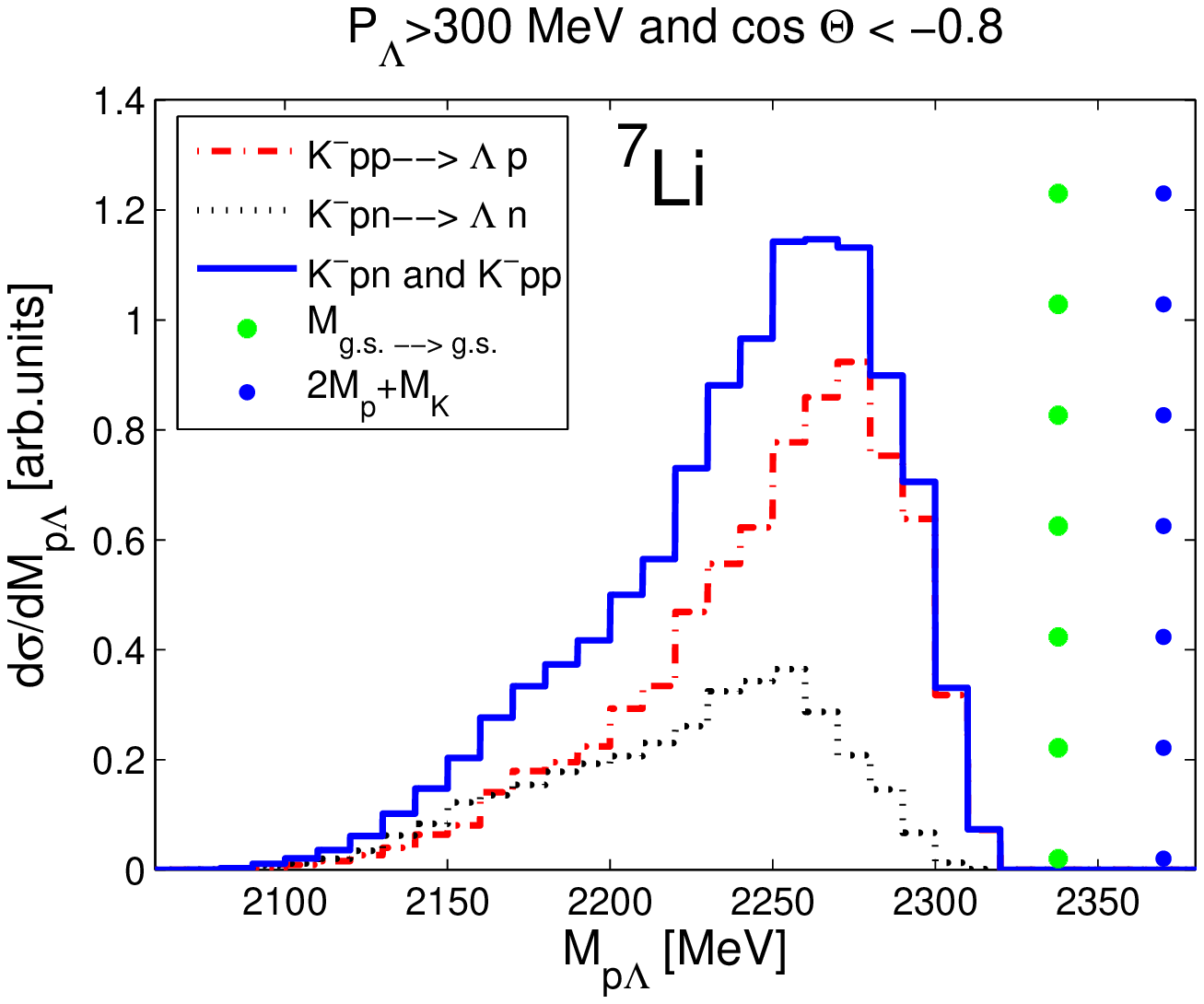}
\end{center}
\vspace{-0.0cm} \caption{(Color online) Invariant mass of $\Lambda p$
distribution for $K^-$ absorption in  $^{7}$Li and allowing up to three collisions of the outgoing particles
with the daughter nucleus.} \label{li7}
\vspace{-0.0cm}
\end{figure}

\begin{figure}
\vspace{-0.0cm}
\begin{center}
\includegraphics[width=8.6cm]{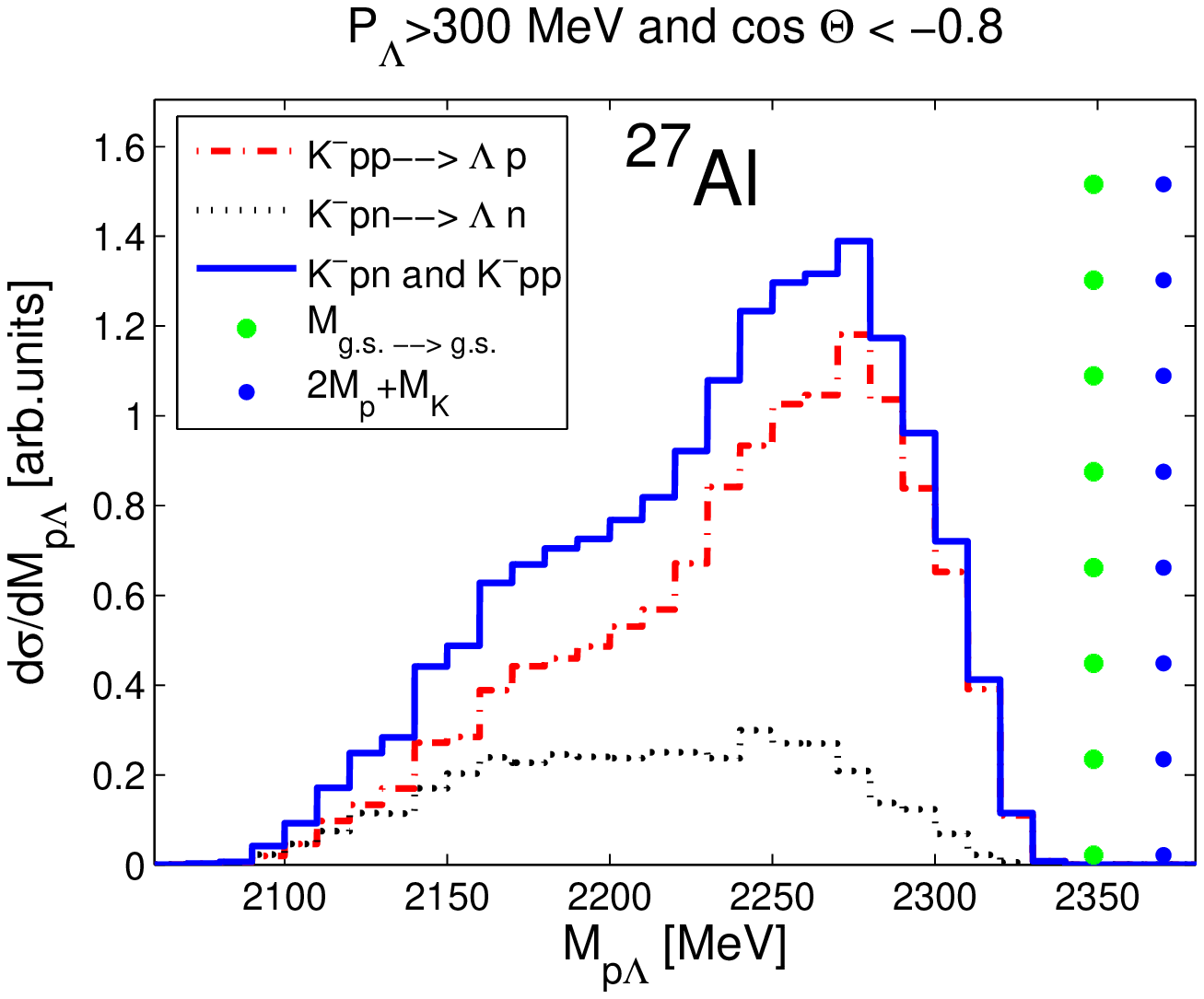}
\end{center}
\vspace{-0.0cm} \caption{(Color online) Invariant mass of $\Lambda p$
distribution for $K^-$ absorption in $^{27}$Al and allowing up to four collisions of the outgoing particles
with the daughter nucleus.} \label{al27}
\vspace{-0.0cm}
\end{figure}

\begin{figure}
\vspace{-0.0cm}
\begin{center}
\includegraphics[width=8.6cm]{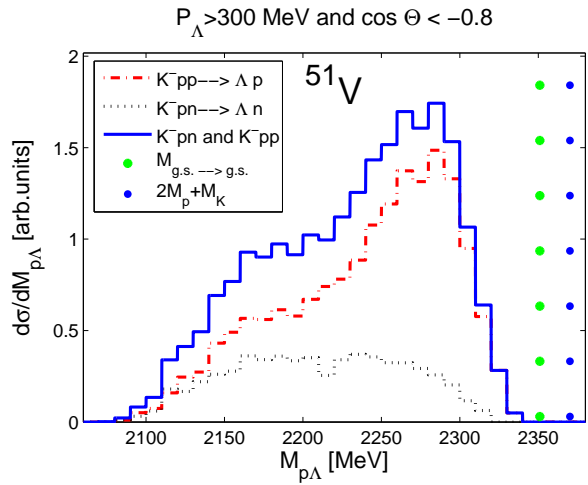}
\end{center}
\vspace{-0.0cm} \caption{(Color online) Invariant mass of $\Lambda p$
distribution for $K^-$ absorption in $^{51}$V and allowing up to five collisions of the outgoing particles
with the daughter nucleus.} \label{v51}
\vspace{-0.0cm}
\end{figure}

In Figs.~\ref{li6}--\ref{v51} we present the calculated
$\Lambda p$ spectra for the other nuclei: $^6$Li, $^7$Li,
$^{27}$Al and $^{51}$V. It is interesting to note that the width
of the distribution
gets broader with the size of the nucleus, while the peak remains in the same
location, consistently to what one expects for the behavior of a quasielastic
peak. We also observe 
that the spectra of heavy nuclei develop a secondary peak at lower invariant masses
due to the larger amount of re-scattering processes as the particles 
move out of the nucleus. 
The spectrum presented in Fig. 3 of Ref.~\cite{finuda} collects data from
the three lighter nuclei. However, to help clarifying the situation, 
it would be desirable 
 to separate the contribution to the invariant mass spectra from
the different nuclei, which might be done in the future.
Let us note in this context that the work of
\cite{Mares:2006vk} shows that the possible interpretation of the FINUDA
peak as a bound state of the $K^-$ with the nucleus, 
not as a $K^- pp$ bound state, 
would unavoidably lead to peaks at different energies for different nuclei.


\begin{figure}
\vspace{-0.0cm}
\begin{center}
\includegraphics[width=8.6cm]{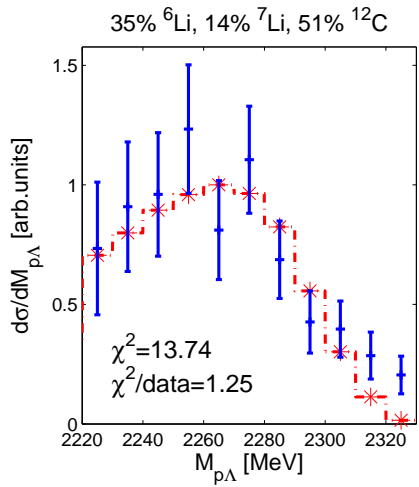}
\end{center}
\vspace{-0.0cm} \caption{(Color online) Invariant mass of $\Lambda p$
distribution for $K^-$ absorption in light nuclei in the following proportion \cite{finuda} including
kaon boost machine corrections \cite{private}: 
51\% $^{12}$C,  35\% $^{6}$Li and 14\% $^{7}$Li. Stars and histogram show 
result of our calculations, experimental points and errorbars are taken from \cite{finuda}.} \label{chi2}
\vspace{-0.0cm}
\end{figure}

Now we want to make a more quantitative comparison of our results with experimental data. We have combined our results for the first three light targets used in the
FINUDA experiment in the same proportion as in the data \cite{finuda} including
kaon
boost machine corrections \cite{private}, namely 51\% $^{12}$C, 35\% $^6$Li and
14\% $^7$Li.
The proper absolute cross sections for the different nuclei (not
those shown in the figures with arbitrary units for each nucleus) are
taken
into account for the weighted nuclear average. The final result, rescaled
to
the arbitrary units of \cite{finuda},
is shown versus experimental data with
corresponding errorbars in Fig. \ref{chi2} for the same energy range as in the
insert
of Fig. 3 in \cite{finuda}. We can see that the agreement is very good, giving
a $\chi^2$ per data point of 1.25. It is important to mention here
that our averaged histogram is dominated by the $^{12}$C component of the
mixture due, not only to the larger amount of protons and the higher
probability of FSI in the bigger nucleus, but also, and mostly,
to the larger overlap with
the kaon wavefunction.
    
In summary, we have seen how the experimental spectrum is naturally
explained in our Monte Carlo simulation 
as a consequence of final state interactions of the
particles produced in nuclear $K^-$ absorption as they leave the nucleus, 
without the need of resorting to exotic mechanisms like the
formation of a $K^- pp$ bound state.

\section{Further mechanisms}

In this section, we explore other possible reactions which could
also produce some strength in the region of energies addressed in
the figures.

We start discussing the mechanism $K^- p p \rightarrow \Sigma^0 p$
followed by $ \Sigma^0 \rightarrow \Lambda \gamma$, mentioned in
\cite{finuda} and discarded there with the claim ``however, the
observed invariant mass distribution is too broad to be attributed
to this process only".

It is easy to check the kinematics by combining relativistically
the momenta of the $p$ from $K^- p p \rightarrow \Sigma^0 p$  and
the $\Lambda$ from $ \Sigma^0 \rightarrow \Lambda \gamma$ and one
obtains that the strength of this channel goes in the $\Lambda p$
invariant mass region between $\sqrt{s}=2200$ MeV and
$\sqrt{s}=2265$ MeV and that all events fulfill the relative angle
condition. With respect to the strength of this mechanism, we note
that, according to the experimental findings of $K^-$ absorption
in $^4$He \cite{55}, there is a fraction of $11.7$ \% for combined
$K^-$ absorption leading to $\Lambda(\Sigma^0)pnn$. This fraction
is split into $2.3$ \%  for $\Sigma^0$ and $9.4$ \% for $\Lambda$
using arguments of isospin symmetry. Given the fact that isospin
symmetry is badly broken at threshold \cite{36}, a different
estimate was done in \cite{osettoki} providing a ratio of
$\Lambda$ to $\Sigma^0$ production of the order of one. For the
estimates we assume an average between these two values, accepting
large uncertainties, and we take $R=\Gamma(\Lambda
p)/\Gamma(\Sigma^0 p)\approx 2$. Assuming now a fraction for the
g.s. formation of the daughter nucleus similar for the two
processes, there would be only about $\sim 5$ \%  of the $K^-$
absorption width going to $\Sigma^0 p$ + g.s. followed by
necessity by the $\Lambda \gamma$ (the main decay mode of the
$\Sigma^0$). Therefore, even if the events fall within the
relevant energy range, the strength of this mechanism is too small
to have a significant effect in the main peak.

 We consider now the $K^-$ absorption
events that go to $\Sigma^0 p$ followed by some collision of the
$p$ or $\Sigma^0$, with final $ \Sigma^0 \rightarrow \Lambda \gamma$
decay as before.
 The fact that the $\Sigma^0$ and $p$ have originally
smaller momentum than the $\Lambda$ and $p$ in the   $K^- p p
\rightarrow \Lambda p$ absorption has as a consequence that fewer
collisions overcome Pauli blocking and bigger collision angles are
required to overcome it. This, together with the $\Lambda$
momentum relative to the $\Sigma^0$, makes very small the fraction
of events that fulfill the experimental condition $\cos
\Theta_{\vec{p}_\Lambda \vec{p}_p}<-0.8$. A simple calculation
shows that assuming $R=2$, only a fraction of about $10$ \%  of
the big peak in Fig. 3 of \cite{finuda} can come from this
process, and these filtered events concentrate around
$\sqrt{s}=2250$ MeV.

Another source of $\Lambda p$ pairs to be investigated is the chain reaction
$K^- N N \rightarrow \Sigma N_1$ followed by $\Sigma N_2 \rightarrow \Lambda p$, with
$N_2$ a nucleon from the Fermi sea. We can have
$$K^- pp \rightarrow \Sigma^+ n\,; \quad \Sigma^+ n \rightarrow \Lambda p$$
\beq
K^- pp \rightarrow \Sigma^0 p\,; \quad \Sigma^0 p \rightarrow \Lambda p
\eeq{eq9}
$$K^- pn \rightarrow \Sigma^0 n\,; \quad \Sigma^0 p \rightarrow \Lambda p$$
According to \cite{55} the combined rate for the three $\Sigma$
formation reactions shown in the left column of Eq.~(\ref{eq9})
would be $3.3$ \% versus 9.9 \%  for $\Lambda p$ formation. If we
take the ratio of $\Lambda$ to $\Sigma^0$ production to be $R=2$,
as discussed earlier, the numbers would be $4.9$ \% versus
$7.8$~\% . Therefore, roughly speaking, we can estimate the rate
of $\Sigma N$ formation as about one half of that of $\Lambda p$.
Next we demand that there is a collision with $\Sigma N
\rightarrow \Lambda p$ conversion. The cross section for this
reaction in the region of interest to us is of the order of $20$
mb \cite{angels_cite}. Using Eq.~(\ref{eq7}) with $\rho=\rho_0/2$,
since in the second part of Eq.~(\ref{eq9}) the $\Lambda p$
conversion occurs only on neutrons or only on protons depending on
the $\Sigma$ charge, one obtains a survival probability
 of the order of $0.63$ in $^{12}$C and, correspondingly, a 37\% probability
 for $\Sigma N \rightarrow \Lambda N$ conversion in the nucleus. All together,
 considering also the larger probability of $NN$ or $\Lambda N$ quasielastic
 collisions we find about one fourth of $\Lambda p$ events coming from this
 source compared to those from $K^- pp \rightarrow \Lambda p$ followed by $\Lambda$
 or $p$ quasielastic collisions. However, when the $\Lambda$ and $p$ momenta in
 their c.m. frame are boosted to the rest frame of the nucleus where the $\Sigma$
 has about $450$ MeV/c, the angle between the $\Lambda$ and the $p$
 span a broad range and the fraction of events that fulfill the condition
 $\cos \Theta_{\vec{p}_\Lambda \vec{p}_p}<-0.8$ is very small. Together with
 the ratio one fourth of the events discussed above these contributions are
 negligible. That small fraction would peak in the region of
 $\sqrt{s}=2170$ MeV, where indeed there is a little bump in Fig. 3 of
 \cite{finuda}.

One can also trace the, even fewer, events recombining the primary
$p$ from $K^- p p \rightarrow \Sigma^0 p$ with the $\Lambda$ from
$\Sigma N \rightarrow \Lambda N$ conversion. The same arguments
also hold in this case and the angle cut constrain leaves a
negligible amount of events from this mechanism to account for the
spectrum of Fig. 3 of \cite{finuda}.

 Note that we have concentrated only on two nucleon absorption events since
 these are the ones leading to highly back to back correlated $\Lambda p$ pairs.
 If one released this condition, other important one-body absorption mechanisms
 would contribute. For instance, considering the Fermi motion of the nucleons, the
 process $K^- N \to \Lambda \pi$ could produce a sizable fraction of $\Lambda$
 hyperons with momenta larger than 300 MeV/c, together with a pion in the 
 $\Delta$ resonance region. Further re-scattering of the
 $\Lambda$ and the pion in the nucleus, or pion absorption, will produce
 relatively slow nucleons generating $\Lambda p$ events at lower invariant
 masses which are, however, uncorrelated in angle. Similar arguments hold for
 processes like $K^- N \to \Sigma \pi$ followed by $\Sigma N \to \Lambda N$
 conversion in the nucleus.

The discussion on other possible sources
of $\Lambda p$ pairs presented in this section 
reinforces our conclusions that the spectrum
shown in Fig. 3 of \cite{finuda} comes essentially from $K^- p N
\rightarrow \Lambda N$ absorption followed by further interaction
of the $N$ or the $\Lambda$ with other nucleons, with a small and
narrow contribution coming from the formation of the g.s. of the
daughter nucleus.

\section{Conclusions}

In summary,  we have seen that the $\Lambda p$ invariant mass
distribution shown in Fig. 3 of \cite{finuda} from  $K^-$
absorption in nuclei is naturally explained in terms of the $K^- p N
\rightarrow \Lambda N$ reaction followed by further interaction of
the $N$ or the $\Lambda$ in the daughter nucleus. The events with
one collision account for the big peak in Fig. 3 of \cite{finuda},
interpreted there as a signal of the formation of the $K^- pp$
bound state. The bump at low invariant masses is accounted by
events with two or more collisions. The small and narrow peak at
higher invariant masses finds here the same interpretation as in
\cite{finuda}, i.e. $K^- p p \rightarrow \Lambda p$ absorption
leaving, however, the daughter nucleus in the ground state.

We have presented results for $^6$Li, $^7$Li, $^{12}$C,
$^{27}$Al and $^{51}$V, all of them measured by the FINUDA experiment,
although the spectrum was only shown for the combined data of the three lighter
nuclei. The width of the distribution increases with the nuclear mass while
the peak stays
in the same location, in accordance with our interpretation of it as
coming from the quasielastic processes. Disentangling the spectrum for each
of the nuclear targets used in the FINUDA experiment would be of particular
relevance because a possible interpretation of the data as evidence of
bound $K^-$ nuclear states would unavoidably
produce the peak at a different energy for each nucleus.

The explanation we have found has obvious experimental
implications. Since the peak, claimed to be a signal for  the $K^-
pp$ bound state, is associated in our interpretation to the $K^- p
N \rightarrow \Lambda N$ absorption followed by final state
interactions of the $N$ or the $\Lambda$ with nucleons in the
daughter nucleus, it should not appear if we produce the $\Lambda
p$ pairs in elementary reactions. On the contrary, according to
the interpretation of \cite{finuda}, if the $K^- pp$ system
couples to $\Lambda p$ with such a large strength, so should the
$\Lambda p$ system couple to the $K^- pp$ state at the $\Lambda p$
invariant mass given by the main peak in the insert of Fig. 3 
of \cite{finuda},
$\sqrt{s}=2260$ MeV. Such invariant masses are easily reachable in
the $pp\rightarrow K^+ \Lambda p$ reaction at low energies.
Although a devoted experiment for the present purpose has not been
done, the present state of the art, with reactions like
$pp\rightarrow K^+ p \pi^- X^+$ being currently considered
\cite{COSY}, indicate that the $pp\rightarrow K^+ \Lambda p$
reaction  is not particularly difficult, and its performance,
looking at the $\Lambda p$ invariant mass, would be most
instructive to further clarify the issue discussed here.

\section{Acknowledgments}
We would like to thank A. Gal for useful comments and for suggesting us to use a
realistic antikaon atomic wave function. We also appreciate useful discussions
with T. Bressani, P. Camerini, N. Grion and S. Piano. 
This work is partly supported by contracts BFM2003-00856 and FIS2005-03142 from
MEC (Spain) and FEDER,
the Generalitat de Catalunya contract 2005SGR-00343,
and the E.U. EURIDICE network contract HPRN-CT-2002-00311.
This research is part of the EU Integrated Infrastructure Initiative
Hadron Physics Project under contract number RII3-CT-2004-506078.

%\appendix
%\section{}
%\label{ap1}

\end{document}